\documentclass[a4paper,fleqn,onecolumn,usenatbib]{mnras}

\usepackage[T1]{fontenc}
\usepackage{ae,aecompl}

\usepackage{graphicx}	
\usepackage{amsmath}	
\usepackage{amssymb}	
\usepackage{graphics}

\newcommand\bnabla{{\bmath\nabla}}
\newcommand\bcdot{{\bmath\cdot}}

\newcommand\bme{{\bmath e}}

\newcommand\bml{{\bmath l}}

\newcommand\bmT{{\bmath T}}

\renewcommand\bmu{{\bmath u}}
\newcommand\bmv{{\bmath v}}

\newcommand\bfT{\mathbf{T}}

\newcommand\half{{\textstyle\frac{1}{2}}}
\newcommand\third{{\textstyle\frac{1}{3}}}
\newcommand\twothirds{{\textstyle\frac{2}{3}}}

\newcommand\rmb{\mathrm{b}}

\newcommand\rmd{\mathrm{d}}
\newcommand\rme{\mathrm{e}}

\newcommand\rmh{\mathrm{h}}
\newcommand\rmi{\mathrm{i}}

\newcommand\rms{\mathrm{s}}
\newcommand\rmv{\mathrm{v}}

\newcommand\rmD{\mathrm{D}}

\newcommand\f{\frac}
\newcommand\p{\partial}
\newcommand\cst{\mathrm{constant}}

\title[Local model of warped discs]
{Hydrodynamics of warps in the local model of astrophysical discs}

\author[Gordon I. Ogilvie]
{Gordon I. Ogilvie\thanks{E-mail: gio10@cam.ac.uk}
\\
Department of Applied Mathematics and Theoretical Physics,
University of Cambridge, Centre for Mathematical Sciences,\\
Wilberforce Road, Cambridge CB3 0WA, UK
}

\date{Accepted 2022 April 01. Received 2022 March 26; in original form 2022 February 07}

\pubyear{2022}

\begin{document}
\label{firstpage}
\pagerange{\pageref{firstpage}--\pageref{lastpage}}
\maketitle

\begin{abstract}
  We show how the local approximation of astrophysical discs, which is
  the basis for the well known model of the shearing box, can be used to study many aspects of the dynamics of
  warped discs.  In the local model, inclination of the orbit of a
  test particle with respect to the reference orbit corresponds to a
  vertical oscillation of the particle at the orbital frequency.
  Warping of a disc corresponds to a locally axisymmetric corrugation
  of the midplane of the disc that oscillates vertically at the
  orbital frequency, while evolution of the warp corresponds to a
  modulation of the complex amplitude of the vertical oscillation.  We
  derive a conservation law for this amplitude that is the local
  equivalent of the conservation of angular momentum and therefore governs the evolution of the warp.  For
  lengthscales that are long compared to the vertical scaleheight of the disc, the
  known non-resonant and resonant regimes of warp dynamics, including the
  diffusive and wavelike regimes of Keplerian discs, occur in the
  local model in the same way as in a global view of warped discs.
\end{abstract}




\section{Introduction}

An astrophysical disc is warped when the plane of its orbital motion
varies with distance from the centre. An initially flat disc will
become warped if it interacts with a companion on a misaligned orbit
\citep{1995MNRAS.274..987P,2013MNRAS.431.1320X} or with a central object with a misaligned
spin axis
\citep{1975ApJ...195L..65B,1999ApJ...524.1030L}. Instabilities can
also generate warps spontaneously in initially aligned systems subject to tidal \citep{1992ApJ...398..525L}, radiation \citep{1996MNRAS.281..357P} or magnetic \citep{1999ApJ...524.1030L} forces. The
existence of warped or misaligned discs has been confirmed by
observations of galactic nuclei (e.g.\ NGC 4258;
\citealt{1995Natur.373..127M}), interacting binary stars (e.g.\ Her
X-1; \citealt{1976ApJ...209..562G}) and, increasingly, young stars
\citep{2019Natur.565..206S,2019MNRAS.486L..58C,2022A&A...658A.183B}.  The Milky Way
itself has also been found to be warped \citep[][and references
therein]{2019Sci...365..478S}.

The hydrodynamics of warped gaseous discs has been studied since the
1970s \citep{1975ApJ...195L..65B}. A significant body of work has been
directed at deriving an evolutionary equation for the tilt vector
$\bml(r,t)$ (a unit vector normal to the local orbital plane of the disc at radius $r$
and time $t$) and, therefore, the shape of the disc
\citep{1978ApJ...226..253P,1983MNRAS.202.1181P,1992MNRAS.258..811P,1995ApJ...438..841P,1999MNRAS.304..557O,2006MNRAS.365..977O}. This
evolution is controlled by the conservation of angular momentum and
therefore involves a calculation of the internal torque in the
disc. In Keplerian discs, the transmission of the warp is relatively
fast because the coincidence of the orbital and epicyclic frequencies
leads to a resonant amplification of the internal flows and associated
torques. In a sufficiently viscous disc, the warp evolves diffusively
on a timescale that is typically much shorter than the viscous
timescale, while in an inviscid disc the warp propagates as a bending
wave at a significant fraction of the sound speed.

Numerical simulations of warped discs have been carried out since the
1990s \citep{1996MNRAS.282..597L,1999MNRAS.309..929N}, mainly using
the SPH method but also, increasingly, with grid-based methods
\citep[e.g.][]{2019ApJ...878..149H,2019MNRAS.487..550L}.  These
simulations are very demanding because they are global and
three-dimensional, involving a huge range of length- and timescales.
It is extremely challenging at present to attempt to resolve both the
global structure and long-term evolution of a thin, warped disc and
the small-scale physics that may be occurring on scales less than the
vertical scaleheight of the disc.

Partly in order to study this small-scale physics,
\citet{2013MNRAS.433.2403O} introduced a local model, the warped
shearing box. This model generalizes the well known shearing box to
incorporate the oscillatory geometry experienced by an observer
orbiting within a warped disc, which is described by a single
dimensionless parameter, the warp amplitude $|\psi|$. The warped
shearing box has been used successfully to study the propagation of
warps in magnetized discs \citep{2018MNRAS.477.2406P} and the
hydrodynamic instability of warped discs in both linear and nonlinear
regimes \citep{2013MNRAS.433.2420O,2019MNRAS.483.3738P}.

The aim of this paper is to propose and develop an alternative way of
studying the physics of warped discs using a local model. It is based
on a standard shearing box and so can make use of existing numerical
codes. Instead of a fixed warp being imposed on the system by an
oscillatory deformation of the coordinate system, as in the warped
shearing box, the warp is represented as part of the solution and
evolves freely.  Crucial to this approach is
the idea that the warp can be identified, in the local model, with the
modulation of a vertical oscillation of the disc.  We develop the
theory underlying this correspondence in Sections~\ref{s:local}
and~\ref{s:symmetries}.  Then, in Section~\ref{s:asymptotic}, we
derive evolutionary equations for the warp in the long-wavelength
limit, showing the detailed correspondence with global asymptotic
theories of warped discs.  In Section~\ref{s:computational} we discuss some relevant computational considerations,
before concluding in Section~\ref{s:conclusion}.

\section{The local model of astrophysical discs}
\label{s:local}

\subsection{Construction of the local model}

Let $(r,\phi,z)$ be cylindrical polar coordinates in an inertial frame
of reference.  We consider a gravitational potential $\Phi(r,z)$ that
has both axial symmetry and reflectional symmetry in the plane $z=0$
and admits a family of stable circular orbits in that plane.  The
angular velocity $\Omega(r)$ of the circular orbit of radius $r$ is
given by
\begin{equation}
  r\Omega^2=\Phi_r(r,0),
\end{equation}
where the subscript denotes a partial derivative, and the angular
frequencies $\kappa(r)$ and $\nu(r)$ of horizontal and vertical
oscillations about that orbit are given by
\begin{equation}
  \kappa^2=\f{1}{r^3}\f{\rmd}{\rmd r}(r^4\Omega^2)=4\Omega^2+2r\Omega\f{\rmd\Omega}{\rmd r}=\Phi_{rr}(r,0)+\f{3}{r}\Phi_r(r,0),\qquad
  \nu^2=\Phi_{zz}(r,0).
\end{equation}
We assume that $\Omega^2$, $\kappa^2$ and $\nu^2$ are positive so that
circular orbits exist and are stable.  For any spherically symmetric
potential (associated with a central force), $\nu=\Omega$.  In the
important special case of a point-mass potential
$\Phi=-GM(r^2+z^2)^{-1/2}$, we have
$\kappa=\nu=\Omega=(GM/r^3)^{1/2}$.

The local model is based on a reference point that follows a selected
circular orbit of radius $r_0$ and angular velocity
$\Omega_0=\Omega(r_0)$.  We set up a Cartesian coordinate system
$(x,y,z)$ with origin at the reference point and axes pointing in the
radial ($x$), azimuthal ($y$) and vertical ($z$) directions.  The
coordinate system therefore rotates with angular velocity $\Omega_0$.
The equation of motion of a test particle in the local approximation
is
\begin{equation}
  \ddot x-2\Omega_0\dot y=2q_0\Omega_0^2x,\qquad
  \ddot y+2\Omega_0\dot x=0,\qquad
  \ddot z=-\nu_0^2z,
\label{xdd}
\end{equation}
with $\nu_0=\nu(r_0)$ and $q_0=q(r_0)$, where
\begin{equation}
  q=-\f{\rmd\ln\Omega}{\rmd\ln r}
\end{equation}
is the dimensionless orbital shear rate.
The terms in equation~(\ref{xdd}) proportional to $2\Omega_0$ come
from the Coriolis force, while the terms proportional to $x$ or $z$
come from the expansion of the sum of the gravitational and
centrifugal forces to first order about the reference point.  The
epicyclic frequency is
\begin{equation}
  \kappa_0=\sqrt{2(2-q_0)}\Omega_0=\kappa(r_0).
\end{equation}
Henceforth we drop the subscript $0$ on $\Omega_0$, $q_0$, $\kappa_0$
and $\nu_0$, so that $\Omega$, $q$, $\kappa$ and $\nu$ are regarded as
constants and correspond to the values of those quantities on the
reference orbit.

The general solution of equations~(\ref{xdd}) is
\begin{equation}
  x=x_0+\mathrm{Re}\left(X\,\rme^{-\rmi\kappa t}\right),\qquad
  y=y_0-q\Omega x_0t+\mathrm{Re}\left(\f{2\Omega}{\rmi\kappa}X\,\rme^{-\rmi\kappa t}\right),\qquad
  z=\mathrm{Re}\left(Z\,\rme^{-\rmi\nu t}\right),
\label{solution}
\end{equation}
where $x_0$ and $y_0$ are real constants and $X$ and $Z$ are complex
constants.  The horizontal part of the solution involves an elliptical
oscillation at the epicyclic frequency around a guiding centre.  The
guiding centre has a fixed radial location and drifts uniformly in the
azimuthal direction if $x_0\ne0$.  Therefore the guiding centre
follows the local representation of a circular orbit, the azimuthal
drift being a consequence of the orbital shear.  The vertical part of
the solution is just a harmonic oscillation at the vertical frequency.

In the case of a spherically symmetric potential, for which
$\nu=\Omega$, the vertical oscillation of a particle in the local
model can be identified with the inclination of the particle's orbit
with respect to the reference orbit.  If we consider a circular orbit
of radius $r$ and angular velocity $\Omega$ in an inclined plane with
unit normal vector $\bml$, then the vertical coordinate oscillates
harmonically in time, such that
$z=\mathrm{Re}\left(Z\,\rme^{-\rmi\Omega t}\right)$ with
$Z=-r(l_x+\rmi l_y)$, where (in this paragraph only) $(x,y,z)$ are
Cartesian coordinates in an inertial frame with origin at the centre
of the potential.  Therefore $-Z/r$ can be identified with the complex
tilt variable $W=l_x+\rmi l_y$ used in numerous previous studies of
warped discs \citep[e.g.][]{1981ApJ...247..677H,1985MNRAS.213..435K}.

For a fluid disc, we replace the equation of motion (\ref{xdd}) with
the equivalent version for a continuous medium,
\begin{align}
  &\rmD u_x-2\Omega u_y=2q\Omega^2x+\f{1}{\rho}(\p_xT_{xx}+\p_yT_{xy}+\p_zT_{xz}),\label{dux}\\
  &\rmD u_y+2\Omega u_x=\f{1}{\rho}(\p_xT_{yx}+\p_yT_{yy}+\p_zT_{yz}),\\
  &\rmD u_z=-\nu^2z+\f{1}{\rho}(\p_xT_{zx}+\p_yT_{zy}+\p_zT_{zz}),
\end{align}
where $\bmu$ is the fluid velocity,
\begin{equation}
  \rmD=\p_t+u_x\p_x+u_y\p_y+u_z\p_z
\end{equation}
is the Lagrangian time-derivative following the fluid motion, $\rho$
is the mass density and $\bfT$ is the stress tensor.  We also require
the equation of mass conservation,
\begin{equation}
  \rmD\rho=-\rho(\p_xu_x+\p_yu_y+\p_zu_z).
\label{drhou}
\end{equation}
The divergence of the stress tensor gives the force per unit volume
resulting from momentum transport within the fluid.  The stress tensor
could include a number of effects such as fluid pressure, viscosity,
magnetic fields, self-gravity, radiation forces, turbulence, etc.  To
maintain generality we do not at this stage write down a constitutive
or evolutionary equation for the stress, which would be required in
order to close the system of equations.

It can be useful to separate the fluid velocity $\bmu$ into a part due
to the orbital shear, $-q\Omega x\,\bme_y$, and a residual velocity
$\bmv$:
\begin{equation}
  u_x=v_x,\qquad
  u_y=-q\Omega x+v_y,\qquad
  u_z=v_z.
\end{equation}
In terms of $\bmv$, the governing equations read
\begin{align}
  &\rmD v_x-2\Omega v_y=\f{1}{\rho}(\p_xT_{xx}+\p_yT_{xy}+\p_zT_{xz}),\\
  &\rmD v_y+(2-q)\Omega v_x=\f{1}{\rho}(\p_xT_{yx}+\p_yT_{yy}+\p_zT_{yz}),\\
  &\rmD v_z=-\nu^2z+\f{1}{\rho}(\p_xT_{zx}+\p_yT_{zy}+\p_zT_{zz}),\\
  &\rmD\rho=-\rho(\p_xv_x+\p_yv_y+\p_zv_z),
\end{align}
with
\begin{equation}
  \rmD=(\p_t-q\Omega x\p_y)+v_x\p_x+v_y\p_y+v_z\p_z.
\end{equation}
These equations are horizontally homogeneous in the sense that their
coefficients do not depend on $x$ or $y$, except for the appearance of
$x\p_y$ in the operator $\rmD$.  This dependence can be removed either
by considering `locally axisymmetric' solutions that are independent
of $y$, as we do in this paper, or by adopting a shearing coordinate
system that follows the orbital shear, although in that case an
explicit time-dependence appears in the equations.

The well known model of the shearing box \citep{1995ApJ...440..742H}
considers these equations in a cuboid together with boundary
conditions on $\bmv$, $\rho$, etc., that are periodic in $y$ and
shearing-periodic in $x$ (i.e.\ periodic in shearing coordinates).
In this paper we consider the equations of the local model without
necessarily applying the boundary conditions of the shearing box.

\subsection{Axisymmetric linear waves in the local model}

\begin{figure}
\centerline{\includegraphics[height=7.5cm]{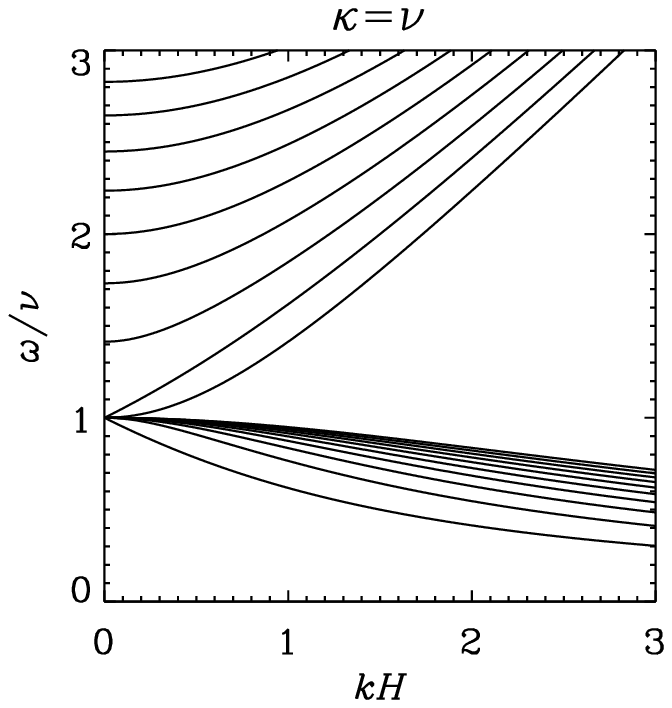}\includegraphics[height=7.5cm]{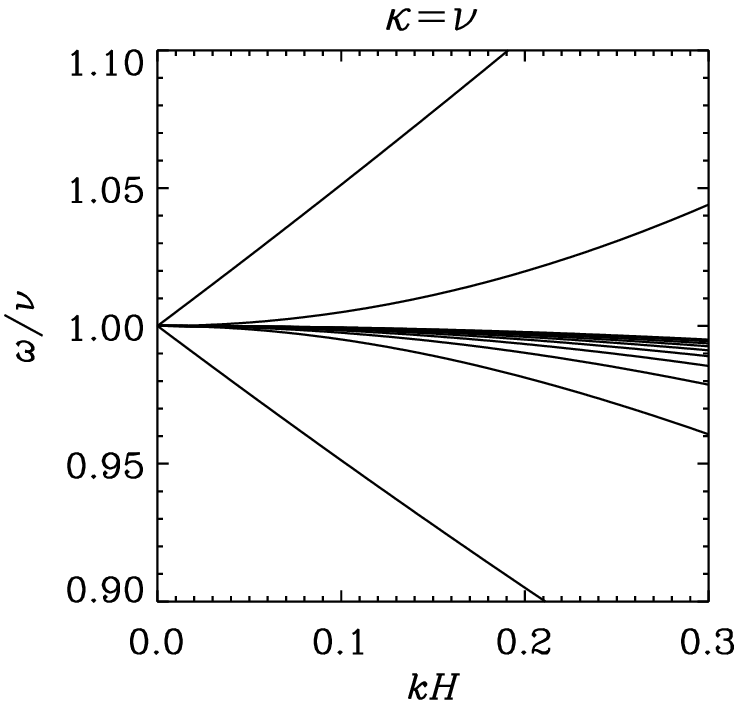}}
\centerline{\includegraphics[height=7.5cm]{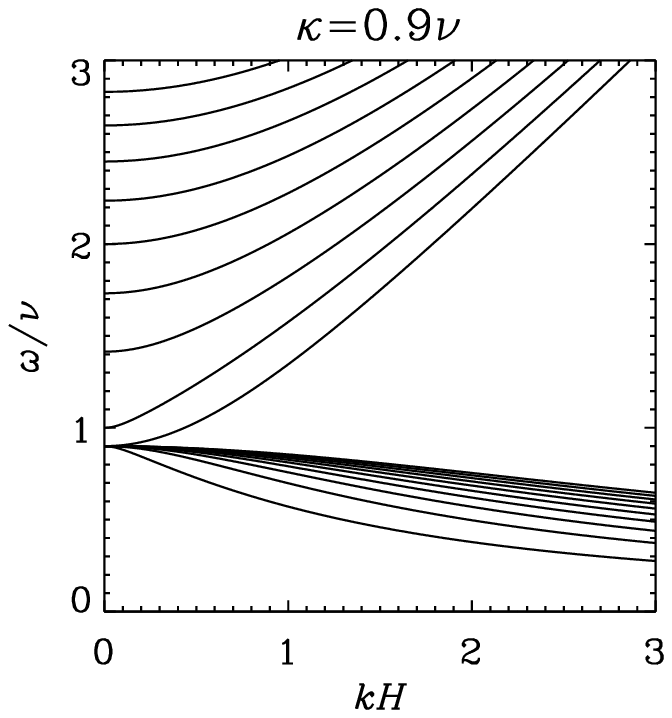}\includegraphics{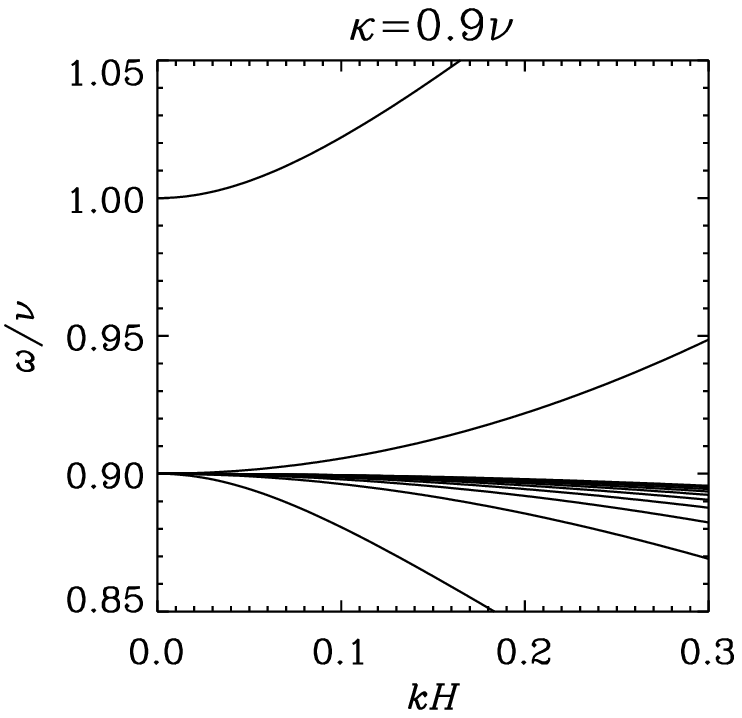}}
\centerline{\includegraphics[height=7.5cm]{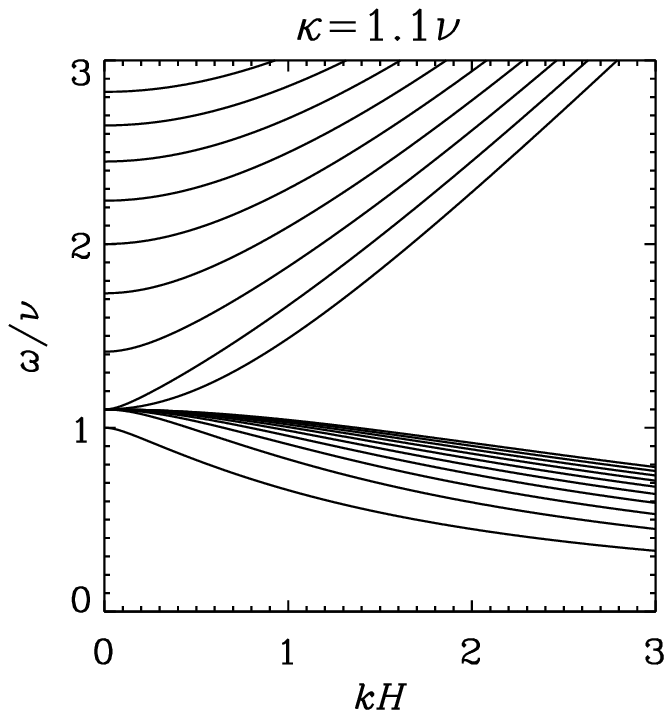}\includegraphics[height=7.5cm]{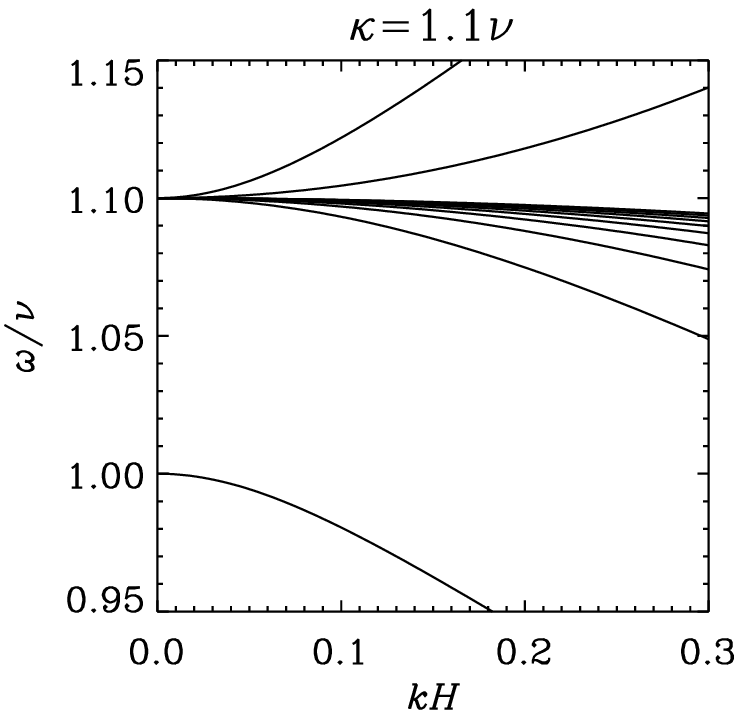}}
\caption{Linear dispersion relations for axisymmetric waves in an
  isothermal disc in the local approximation. Modes up to $n=10$ are
  plotted. The right panels zoom in to the region relevant for
  long-wavelength bending modes.}
\label{f:waves}
\end{figure}

Locally axisymmetric solutions that are independent of $y$ satisfy the
equations
\begin{align}
  &\rmD v_x-2\Omega v_y=\f{1}{\rho}(\p_xT_{xx}+\p_zT_{xz}),\label{ddvx}\\
  &\rmD v_y+(2-q)\Omega v_x=\f{1}{\rho}(\p_xT_{yx}+\p_zT_{yz}),\label{ddvy}\\
  &\rmD v_z=-\nu^2z+\f{1}{\rho}(\p_xT_{zx}+\p_zT_{zz}),\label{ddvz}\\
  &\rmD\rho=-\rho(\p_xv_x+\p_zv_z),\label{ddrho}
\end{align}
where
\begin{equation}
  \rmD=\p_t+v_x\p_x+v_z\p_z.
\end{equation}
We consider here the simplest situation of an ideal fluid that is an
isothermal gas, so that all components of $\bfT$ vanish except for an
isotropic pressure that is proportional to the density. Thus
\begin{equation}
  T_{xx}=T_{yy}=T_{zz}=-c_\rms^2\rho,
\label{isothermal}
\end{equation}
where $c_\rms=\cst$ is the isothermal sound speed.

The simplest solution of the equations is the vertically hydrostatic
basic state
\begin{equation}
  v_x=v_y=v_z=0,\qquad
  \rho=\rho_0\exp\left(-\f{z^2}{2H^2}\right),\qquad
  H=\f{c_\rms}{\nu},
\end{equation}
in which there is no departure from circular orbital motion, and the
density and pressure are Gaussian functions of $z$ with scaleheight
(or standard deviation) $H$.

Linear wave modes on this background that depend on $x$ and $t$
through the factor $\exp(\rmi kx-\rmi\omega t)$ have the following
structure \citep{1987PASJ...39..457O,1999ApJ...515..767O}:
\begin{equation}
  v_x\propto v_y\propto\f{\rho'}{\rho}\propto\mathrm{He}_n\left(\f{z}{H}\right),\qquad v_z\propto\mathrm{He}_{n-1}\left(\f{z}{H}\right),\qquad n=0,1,2,\dots,
\end{equation}
involving the Hermite polynomials. The wave's angular frequency $\omega$, radial wavenumber $k$ and vertical mode number $n$ satisfy the dispersion relation
\begin{equation}
  (\omega^2-n\nu^2)(\omega^2-\kappa^2)=c_\rms^2k^2\omega^2.
\end{equation}
We focus on the modes with vertical mode number $n=1$, which are
related to warping or bending disturbances of the disc \citep[e.g.][]{1995ApJ...438..841P}.  The structure
of these modes is such that $v_z$ is independent of $z$, while
$v_x,v_y\propto z$, with
\begin{equation}
  \f{v_x}{v_z}=\left(\f{\omega^2}{\omega^2-\kappa^2}\right)\rmi kz,\qquad
  \f{v_y}{v_x}=\f{(2-q)\Omega}{\rmi\omega}.
\label{vxvz}
\end{equation}
When $k=0$ there is an $n=1$ mode with frequency $\omega=\nu$,
corresponding to a horizontally uniform vertical oscillation of the
disc ($v_z=\cst$) at the vertical frequency.  For $0<kH\ll1$ this mode
becomes a long-wavelength bending mode with a frequency slightly
different from $\nu$.

Provided that $\kappa\ne\nu$, which we refer to as the
\textit{non-resonant case}, the $n=1$ bending mode has the
long-wavelength expansion
\begin{equation}
  \omega=\nu\left\{1+\f{\nu^2}{2(\nu^2-\kappa^2)}(kH)^2+O\left[(kH)^4\right]\right\}.
\end{equation}
The group velocity $\rmd\omega/\rmd k$ is proportional to $k$ for
sufficiently small values of $kH$, indicating that the waves are dispersive. We see from
equation~(\ref{vxvz}) that this mode involves some horizontal motion
proportional to $z$ in addition to the vertical motion independent of
$z$. The horizontal motion is forced by radial pressure gradients
associated with the radial variation of the vertical oscillation.

In the \textit{resonant case} $\kappa=\nu$, which includes the
important special case of a point-mass potential, a different
expansion is required.  We have instead
\begin{equation}
  \omega=\nu\left\{1\pm\f{kH}{2}+O\left[(kH)^2\right]\right\},
\label{resonant}
\end{equation}
corresponding to a pair of waves with constant group velocities $\pm
c_\rms/2$ in the limit $kH\ll1$.

Examples of the dispersion relations in resonant and non-resonant
cases are plotted in Fig.~\ref{f:waves}.  The right-hand panels zoom
in to the region where $kH\ll1$ and $\omega$ is close to $\nu$.  We
can think of the solutions on the relevant branches as long-wavelength
bending waves in which, to a first approximation, each column of the
disc undergoes a harmonic oscillation at the vertical frequency; this
oscillation is modulated on a longer timescale, resulting in a
frequency slightly different from $\nu$, because of the finite
horizontal wavelength of the corrugation and the communication of
vertical momentum between neighbouring columns of the disc.

\subsection{Relation to a warped disc}

\begin{figure}
\centerline{\includegraphics[height=13cm]{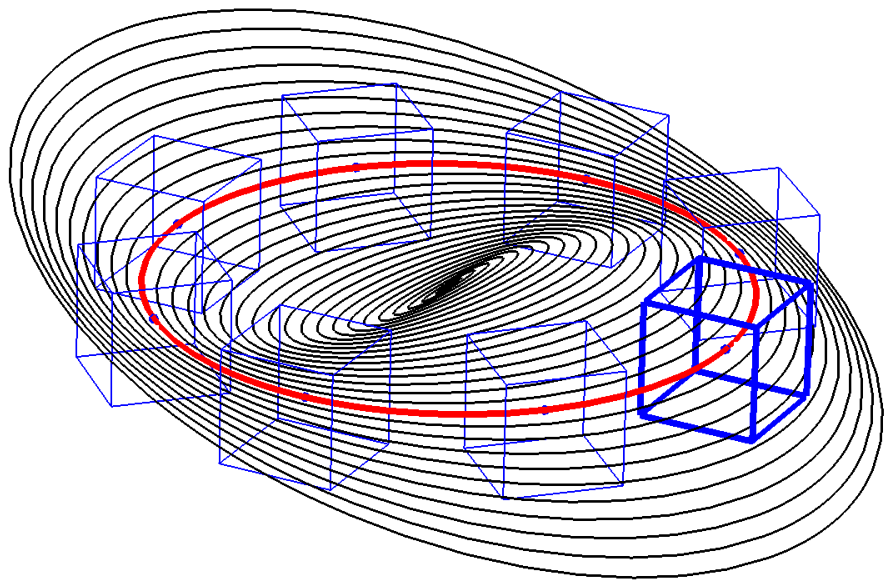}}
\vskip-1cm
\centerline{\includegraphics[height=9cm]{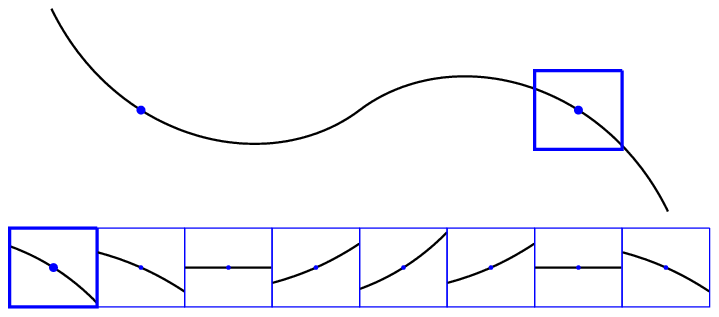}}
\caption{Appearance of a warped disc in the local model. The upper
  panel shows an example of a warped disc in which a reference orbit
  (red circle) is selected and used to construct a local model.  The
  local frame (blue cube) is illustrated at eight equally spaced
  orbital phases.  The lower panel presents a side-on view of the
  warped disc at the initial orbital phase; below this, a succession
  of boxes shows the appearance of the warped disc in the local frame
  at eight equally spaced orbital phases.  To the extent that the warp
  is stationary in the non-rotating frame, each part of the disc
  oscillates vertically at the orbital frequency in the local frame.
  The illustrated warp is untwisted, so the phase of the vertical
  oscillation does not vary with radial location.}
\label{f:warp}
\end{figure}

It might seem obvious that a warped disc is a global, large-scale and
non-axisymmetric phenomenon.  The idea that the physics of warped
discs can be captured in a local, axisymmetric model therefore
requires some explanation.  Fig.~\ref{f:warp} illustrates the
construction of a local model around a reference point in a circular
orbit in a warped disc.  The orbital motion at neighbouring radial
locations in the disc is slightly misaligned with the reference orbit.
The relative motion appears, in the local model, as a vertical
oscillation of the disc with an amplitude and phase that depend on the
radial location.  To the extent that the warp is stationary in the
non-rotating frame, each part of the disc oscillates vertically at the
orbital frequency in the local frame; any slow evolution of the warp
in the non-rotating frame would correspond to a temporal modulation of
the local vertical oscillation.  The vertical oscillation is also
locally axisymmetric (independent of $y$), to the extent that the
azimuthal length of the box is small compared to the circumference of
the disc.

\section{Symmetries and conservation laws of the local model}
\label{s:symmetries}

\subsection{Particle dynamics}

The equation of motion~(\ref{xdd}) of a test particle in the local
approximation is invariant under the addition of a uniform vertical
oscillation of angular frequency $\nu$ and of arbitrary amplitude and
phase.  Thus if $z$ is replaced by $z+\zeta$, where $\zeta(t)$
satisfies $\ddot\zeta+\nu^2\zeta=0$, and all other variables remain
unchanged, the equations are invariant.  In the case of a spherically
symmetric gravitational potential for which $\nu=\Omega$, the
transformation $z\mapsto z+\zeta$ corresponds to a redefinition of the
orientation of the reference orbit about which the local model is
constructed.

Associated with this continuous symmetry is a conservation law for the complex quantity
\begin{equation}
  Z=\left(z+\f{\rmi\dot z}{\nu}\right)E_\nu,\qquad
  E_\nu=\rme^{\rmi\nu t},
\end{equation}
which represents both the amplitude and phase of the vertical
oscillation and agrees with the quantity $Z$ in the general
solution~(\ref{solution}). This conservation law can be derived from
Noether's Theorem, and is in fact just one of several conservation
laws that hold in the local model when both horizontal and vertical motion are considered.

\subsection{Continuum mechanics}

In a similar way, the differential equations
(\ref{dux})--(\ref{drhou}) and boundary conditions of a continuous
medium in the local model are invariant under the addition of a
uniform vertical oscillation of angular frequency $\nu$ and of
arbitrary amplitude and phase.  Thus if $z$ and $u_z$ are replaced by
$z+\zeta$ and $u_z+\dot\zeta$, where $\zeta(t)$ satisfies
$\ddot\zeta+\nu^2\zeta=0$, and all other variables remain unchanged,
the equations are invariant.

Associated with this continuous symmetry is a conservation law 
for the complex quantity
\begin{equation}
  Z=\left(z+\f{\rmi u_z}{\nu}\right)E_\nu,\qquad
  E_\nu=\rme^{\rmi\nu t},
\end{equation}
which is clearly analogous to the complex amplitude of the vertical
oscillation of a test particle, as considered above.  Indeed, starting from equations
(\ref{dux})--(\ref{drhou}) we can obtain the equation
\begin{equation}
  \p_t(\rho Z)+\p_x\left(\rho Zu_x-\f{\rmi T_{zx}E_\nu}{\nu}\right)+\p_y\left(\rho Zu_y-\f{\rmi T_{zy}E_\nu}{\nu}\right)+\p_z\left(\rho Zu_z-\f{\rmi T_{zz}E_\nu}{\nu}\right)=0,
\label{z}
\end{equation}
which is in conservative form.

We can relate the conservation of $Z$ to the conservation of angular
momentum.  Let us consider a spherically symmetric gravitational
potential and temporarily employ a Cartesian coordinate system in an
inertial frame with origin at the centre of the potential.  Then the
conservative form of the angular momentum equation is
\begin{equation}
  \p_t(\rho\epsilon_{ijk}x_ju_k)+\p_l(\rho\epsilon_{ijk}x_ju_ku_l-\epsilon_{ijk}x_jT_{kl})=0,
\end{equation}
provided that the stress tensor is symmetric.  The horizontal
components of this equation are
\begin{align}
  &\p_t[\rho(yu_z-zu_y)]+\p_l[\rho(yu_z-zu_y)u_l-yT_{zl}+zT_{yl}]=0,\\
  &\p_t[\rho(zu_x-xu_z)]+\p_l[\rho(zu_x-xu_z)u_l-zT_{xl}+xT_{zl}]=0.
\end{align}
Combining these components in the complex linear combination $x+\rmi
y$, we obtain
\begin{equation}
  \p_t\left\{-\rmi\rho[(x+\rmi y)u_z-z(u_x+\rmi u_y)]\right\}+\p_l\left\{-\rmi\rho[(x+\rmi y)u_z-z(u_x+\rmi u_y)]u_l+\rmi(x+\rmi y)T_{zl}-\rmi z(T_{xl}+\rmi T_{yl})\right\}=0.
\end{equation}
In cylindrical polar coordinates this reads
\begin{equation}
  \p_t\left\{-\rmi\rho[ru_z-z(u_r+\rmi u_\phi)]\,\rme^{\rmi\phi}\right\}+\bnabla\bcdot\left\{-\rmi\rho[ru_z-z(u_r+\rmi u_\phi)]\,\rme^{\rmi\phi}\bmu+\rmi[r\bmT_z-z(\bmT_r+\rmi\bmT_\phi)]\,\rme^{\rmi\phi}\right\}=0,
\end{equation}
where $\bmT_i=T_{ij}\,\bme_j$ is (minus) the flux density of the $i$-component
of momentum.  If we now adopt the local approximation by selecting a
reference orbit of radius $r_0$, going into a frame that rotates with
that orbit and constructing a local Cartesian coordinate system, then
equation~(\ref{z}) (multiplied by the constant $-r_0\Omega_0$) emerges
as the leading approximation to this angular-momentum equation.  To
see this we must note that $|z|\ll r\approx r_0$ and
$|u_r|\ll u_\phi\approx r_0\Omega_0$. The phase factor
$\rme^{\rmi\phi}$ translates into $\rme^{\rmi\Omega_0t}$ in the
leading approximation, and we note that $\nu=\Omega$ for a spherically
symmetric potential as considered here.
Therefore the conservation of $Z$ in the local model is directly
related to the conservation of horizontal angular momentum in the
global description.  The reason for this association is that, by
assumption, the angular momentum of the fluid is dominated by its
orbital motion. The horizontal angular momentum derives from the
inclination of the orbital motion, which appears in the local model as
a vertical oscillation.

\section{Long-wavelength corrugations and connection with warped discs}
\label{s:asymptotic}

The main aim of this section is to demonstrate a correspondence
between the dynamics of long-wavelength corrugations in the local
model and the known theories of warped discs that were derived from
asymptotic analysis in spherical geometry.  We first introduce a
warped coordinate system that follows the corrugation, before making
separate asymptotic analyses of the non-resonant case (allowing for
nonlinearity and viscosity) and the resonant case.

\subsection{Warped coordinates}

We return to equations (\ref{ddvx})--(\ref{ddrho}) governing locally
axisymmetric solutions in the local model.
To study the dynamics of a $y$-independent corrugation, which, as we
have seen, is the local representation of a warp, we introduce the
coordinate transformation
\begin{equation}
  z'=z-\zeta(x,t),\qquad
\end{equation}
where $\zeta(x,t)$ describes the corrugation whose dynamics is to be
determined.  The warped midplane corresponds to $z=\zeta(x,t)$ or
$z'=0$. The chain rule gives
\begin{equation}
  \p_x=\p_x'-\zeta_x\p_z',\qquad
  \p_z=\p_z',\qquad
  \p_t=\p_t'-\zeta_t\p_z',
\end{equation}
where $\zeta_x$ and $\zeta_t$ are the partial derivatives of
$\zeta(x,t)$ with respect to $x$ and $t$, and our shorthand notation
for partial derivatives is
\begin{equation}
  \p_x=\left(\f{\p}{\p x}\right)_{z,t},\qquad
  \p_z=\left(\f{\p}{\p z}\right)_{x,t},\qquad
  \p_t=\left(\f{\p}{\p t}\right)_{x,z},\qquad
  \p_x'=\left(\f{\p}{\p x}\right)_{z',t},\qquad
  \p_z'=\left(\f{\p}{\p z'}\right)_{x,t},\qquad
  \p_t'=\left(\f{\p}{\p t}\right)_{x,z'}.
\end{equation}
It can be helpful to introduce the relative vertical velocity
\begin{equation}
  v_z'=v_z-(\zeta_t+v_x\zeta_x)=v_z-\rmD\zeta,
\end{equation}
which differs from the absolute vertical velocity $v_z$ by excluding the vertical velocity $\rmD\zeta$ associated with the
time-dependent corrugation. The Lagrangian derivative is then
\begin{equation}
  \rmD=\p_t'+v_x\p_x'+v_z'\p_z',
\end{equation}
and the velocity divergence is
\begin{equation}
  \Delta=\p_x'v_x+\p_z'v_z'.
\end{equation}
The basic equations become
\begin{align}
  &\rmD v_x-2\Omega v_y=\f{1}{\rho}[\p_x'T_{xx}+\p_z'(T_{xz}-\zeta_xT_{xx})],\label{dvx}\\
  &\rmD v_y+(2-q)\Omega v_x=\f{1}{\rho}[\p_x'T_{yx}+\p_z'(T_{yz}-\zeta_xT_{yx})],\label{dvy}\\
  &\rmD v_z=-\nu^2z+\f{1}{\rho}[\p_x'T_{zx}+\p_z'(T_{zz}-\zeta_xT_{zx})],\label{dvz}\\
  &\rmD\rho=-\rho\Delta.\label{drho}
\end{align}
An alternative form of equation~(\ref{dvz}) that is more useful for
some purposes is
\begin{equation}
  \rmD v_z'=-\nu^2z'-(\rmD^2+\nu^2)\zeta+\f{1}{\rho}[\p_x'T_{zx}+\p_z'(T_{zz}-\zeta_xT_{zx})].
\end{equation}
The conservative forms of these equations are
\begin{align}
  &\p_t'\rho+\p_x'(\rho v_x)+\p_z'(\rho v_z')=0,\\
  &\p_t'(\rho v_x)+\p_x'(\rho v_xv_x-T_{xx})+\p_z'(\rho v_xv_z'-T_{xz}+\zeta_xT_{xx})=2\Omega\rho v_y,\\
  &\p_t'(\rho v_y)+\p_x'(\rho v_yv_x-T_{yx})+\p_z'(\rho v_yv_z-T_{yz}+\zeta_xT_{yx})=-(2-q)\Omega\rho v_x,\\
  &\p_t'(\rho Z')+\p_x'\left(\rho Z'v_x-\f{\rmi T_{zx}E_\nu}{\nu}\right)+\p_z'\left[\rho Z'v_z'-\f{\rmi(T_{zz}-\zeta_xT_{zx})E_\nu}{\nu}\right]=-\f{\rmi\rho}{\nu}E_\nu(\rmD^2+\nu^2)\zeta,
\end{align}
in which the right-hand sides represent source terms, and where
\begin{equation}
  Z'=\left(z'+\f{\rmi v_z'}{\nu}\right)E_\nu,\qquad
  E_\nu=\rme^{\rmi\nu t}.
\end{equation}
We integrate these conservative forms with respect to $z'$ over the
full vertical extent of the disc and assume that there are no net
fluxes of mass or momentum through the vertical boundaries.  (This
important assumption should be reconsidered if the disc has a
significant mass outflow, is self-gravitating, or has an external
magnetic field.)  Thus
\begin{align}
  &\p_t'\int\rho\,\rmd z'+\p_x'\int\rho v_x\,\rmd z'=0,\label{dintrho}\\
  &\p_t'\int\rho v_x\,\rmd z'+\p_x'\int(\rho v_xv_x-T_{xx})\,\rmd z'=2\Omega\int\rho v_y\,\rmd z',\label{dintrhovx}\\
  &\p_t'\int\rho v_y\,\rmd z'+\p_x'\int(\rho v_yv_x-T_{yx})\,\rmd z'=-(2-q)\Omega\int\rho v_x\,\rmd z',\label{dintrhovy}\\
  &\p_t'\int\rho Z'\,\rmd z'+\p_x'\int\left(\rho Z'v_x-\f{\rmi T_{zx}E_\nu}{\nu}\right)\,\rmd z'=-\f{\rmi E_\nu}{\nu}\int\rho(\rmD^2+\nu^2)\zeta\,\rmd z'.\label{dintrhoz}
\end{align}

\subsection{Non-resonant case}
\label{s:non-resonant}

\subsubsection{Asymptotic expansions}

So far the equations are valid for any $y$-independent solution, with
$\zeta(x,t)$ being an arbitrarily specified function.  In order to
deduce the dynamical evolution of the corrugation, we make an
asymptotic analysis of slowly modulated oscillating corrugations using
the method of multiple scales \citep[e.g.][]{BO78}.  Taking the
vertical scaleheight and orbital timescale as the characteristic
scales of length and time, we write the corrugation as
\begin{equation}
  \zeta=\epsilon^{-1}\zeta_0(X,t,T)=\epsilon^{-1}\mathrm{Re}\left[Z_0(X,T)\,\rme^{-\rmi\nu t}\right],
\label{zeta}
\end{equation}
where $\epsilon\ll1$ is a small dimensionless parameter and $X$ and
$T$ are slow space and time coordinates defined by
\begin{equation}
  x=\epsilon^{-1}X,\qquad
  t=\epsilon^{-2}T.
\end{equation}
The meaning of these expressions is as follows.  We are making a
formal separation between the fast orbital timescale (described
by~$t$) and the slow modulatory timescale (described by~$T$).  The
corrugation is of large amplitude ($\epsilon^{-1}$) compared to the
vertical scaleheight.  It consists of a harmonic oscillation on the
orbital timescale, with angular frequency $\nu$, and with an amplitude
and phase (described by the complex amplitude $Z_0$) that vary on a
radial lengthscale that is long ($\epsilon^{-1}$) compared to the
vertical scaleheight and on a timescale that is very slow
($\epsilon^{-2}$) compared to the orbital timescale. The specific scaling of the very slow timescale adopted here, which is related to that used in \citet{1999MNRAS.304..557O}, is designed to capture the evolution of a large-scale warp due to both pressure and viscosity, allowing for the possibility that the viscosity parameter $\alpha$ is $O(1)$ in general.

The quantity $Z_0$ agrees with the complex amplitude $Z$ introduced in
Section~\ref{s:symmetries} in the sense that $Z\sim\epsilon^{-1}Z_0$.
The amplitude of the corrugation may be comparable to the radial
lengthscale on which it varies.  Thus the corrugation gradient
\begin{equation}
  \zeta_x=\zeta_{0X}=\mathrm{Re}\left(Z_{0X}\,\rme^{-\rmi\nu t}\right)
\end{equation}
is of order unity, indicating that the corrugation is nonlinear.
(Subscripts $X$ and $T$ will denote partial derivatives with respect
to the slow variables.) We may write
\begin{equation}
  \zeta_{0X}=\mathrm{Re}\left(-\psi\,\rme^{-\rmi\nu t}\right)=-|\psi|\cos\tau,
\label{zeta0X}
\end{equation}
where $\psi(X,T)=-Z_{0X}$ is the dimensionless complex warp amplitude
used in previous work \citep{1999MNRAS.304..557O}, $|\psi|=|Z_{0X}|$ is
its magnitude and $\tau=\nu t-\arg\psi$ is a phase variable for the
vertical oscillation.

The fluid variables associated with such a corrugation are generally
required to have the following asymptotic expansions in order for the equations to balance:
\begin{align}
  &v_x=v_{x0}(X,z',t,T)+\epsilon v_{x1}(X,z',t,T)+\epsilon^2v_{x2}(X,z',t,T)+\cdots,\\
  &v_y=v_{y0}(X,z',t,T)+\epsilon v_{y1}(X,z',t,T)+\epsilon^2v_{y2}(X,z',t,T)+\cdots,\\
  &v_z'=v_{z0}'(X,z',t,T)+\epsilon v_{z1}'(X,z',t,T)+\epsilon^2v_{z2}'(X,z',t,T)+\cdots,\\
  &\rho=\rho_0(X,z',t,T)+\epsilon\rho_1(X,z',t,T)+\epsilon^2\rho_2(X,z',t,T)+\cdots,\\
  &\bfT=\bfT_0(X,z',t,T)+\epsilon\bfT_1(X,z',t,T)+\epsilon^2\bfT_2(X,z',t,T)+\cdots.
\end{align}
These expressions allow for internal velocities that are comparable to
the sound speed and (generally anisotropic) stresses that are
comparable to the pressure.  Each term depends on $t$ because the
nonlinearity of the corrugation causes all quantities to oscillate on
the orbital timescale at leading order.  Note that the absolute
vertical velocity $v_z=v_z'+\rmD\zeta$ has a different expansion,
\begin{equation}
  v_z=\epsilon^{-1}\zeta_{0t}+v_{z0}(X,z',t,T)+\epsilon v_{z1}(X,z',t,T)+\epsilon^2v_{z2}(X,z',t,T)+\cdots,
\end{equation}
because it includes the large velocity associated with the
time-dependent corrugation.

In the method of multiple timescales, $t$ and $T$ are regarded as
independent variables.  So when the operator $\p_t'$ (in which $x$ and
$z'$ are held constant) acts on any of the above quantities such as
$v_x$, it has the action
\begin{equation}
  \p_t'+\epsilon^2\p_T'=\left(\f{\p}{\p t}\right)_{X,z',T}+\epsilon^2\left(\f{\p}{\p T}\right)_{X,z',t}.
\end{equation}
The Lagrangian time-derivative then has the expansion
\begin{align}
  &\rmD=\rmD_0+\epsilon\rmD_1+\epsilon^2\rmD_2+\cdots,\\
  &\rmD_0=\p_t'+v_{z0}'\p_z',\qquad
  \rmD_1=v_{x0}\p_X'+v_{z1}'\p_z',\qquad
  \rmD_2=\p_T'+v_{x1}\p_X'+v_{z2}'\p_z'.
\end{align}

When these expansions are substituted into the basic equations
(\ref{dvx})--(\ref{drho}) and terms of the same order in $\epsilon$
are compared, we obtain a number of new equations.  The one of lowest
order comes from the vertical component~(\ref{dvz}) of the equation of
motion at $O(\epsilon^{-1})$, which yields
\begin{equation}
  \zeta_{0tt}=-\nu^2\zeta_0.
\end{equation}
This equation is satisfied by our assumption~(\ref{zeta}), which in
fact is the general solution of this equation.  Next, the four
equations at $O(\epsilon^0)$ yield
\begin{align}
  &\rmD_0v_{x0}-2\Omega v_{y0}=\f{1}{\rho_0}\p_z'(T_{xz0}-\zeta_{0X}T_{xx0}),\label{dvx0}\\
  &\rmD_0v_{y0}+(2-q)\Omega v_{x0}=\f{1}{\rho_0}\p_z'(T_{yz0}-\zeta_{0X}T_{yx0}),\label{dvy0}\\
  &v_{x0}\zeta_{0Xt}+\rmD_0v_{z0}=-\nu^2z'+\f{1}{\rho_0}\p_z'(T_{zz0}-\zeta_{0X}T_{zx0}),\label{dvz0}\\
  &\rmD_0\rho_0=-\rho_0\p_z'v_{z0}',\label{drho0}
\end{align}
with
\begin{equation}
  v_{z0}'=v_{z0}-\zeta_{0X}v_{x0}.
\end{equation}
These form a closed system of equations (except for the specification
of the stress tensor) that describe the nonlinear oscillations of the
fluid variables on the orbital timescale in response to the
corrugation. Note that the corrugation appears only through the terms (cf.\ equation~\ref{zeta0X})
\begin{equation}
  \zeta_{0X}=-|\psi|\cos\tau,\qquad
  \zeta_{0Xt}=\nu|\psi|\sin\tau,
\end{equation}
which involve the local warp amplitude $|\psi|$. Furthermore, the
equations do not involve any derivatives with respect to $X$, so they
are local in $X$. They are in fact equivalent to the equations for
horizontally invariant solutions in the warped shearing box
\citep{2013MNRAS.433.2403O} and also to Set~A of the global asymptotic
description of \citet{1999MNRAS.304..557O}. We develop this
correspondence in Section~\ref{s:nonlinear} below.

We assume that the relevant solution of these equations is periodic in
$t$, with the same period $2\pi/\nu$ as the vertical oscillation
associated with the corrugation. (This is the equivalent in the local
model of the assumption in a global model that the disc is stationary
on the orbital timescale and evolves only on a slower timescale.) In a
sufficiently dissipative disc we would expect the solution of these
equations to converge towards such a periodic solution starting from
general initial conditions. In a non-dissipative disc, additional free
oscillation modes could persist unless the initial conditions are
chosen correctly. The laminar oscillatory flows in a warped disc can
be unstable \citep{2000MNRAS.318.1005G,2013MNRAS.433.2420O}, leading to turbulent motion
with a complicated dependence on $x$ and $t$
\citep{2019MNRAS.483.3738P}. If such an instability is present, the
present analysis could be taken to describe the oscillatory mean
flows on the turbulent background, if the turbulent stresses are
represented within $\bfT$.

\subsubsection{Evolutionary equations}

In order to determine the evolution of the corruguation, we do require
some information from higher orders in $\epsilon$.  It is more
convenient for this purpose to use the conservative forms of the
equations.  When the asymptotic expansions are applied to
equation~(\ref{dintrho}) for mass conservation, we obtain, at
$O(\epsilon^0)$, $O(\epsilon^1)$ and $O(\epsilon^2)$,
\begin{align}
  &\p_t'\int\rho_0\,\rmd z'=0,\label{dintrho0}\\
  &\p_t'\int\rho_1\,\rmd z'+\p_X'\int\rho_0v_{x0}\,\rmd z'=0,\\
  &\p_T'\int\rho_0\,\rmd z'+\p_t'\int\rho_2\,\rmd z'+\p_X'\int(\rho_1v_{x0}+\rho_0v_{x1})\,\rmd z'=0.\label{dintrho2}
\end{align}
Equations (\ref{dintrhovx}) and (\ref{dintrhovy}) for horizontal
momentum conservation at $O(\epsilon^0)$ yield
\begin{align}
  &\p_t'\int\rho_0v_{x0}\,\rmd z'=2\Omega\int\rho_0v_{y0}\,\rmd z',\label{dintrhovx0}\\
  &\p_t'\int\rho_0v_{y0}\,\rmd z'=-(2-q)\Omega\int\rho_0v_{x0}\,\rmd z',\label{dintrhovy0}
\end{align}
and at $O(\epsilon^1)$ yield
\begin{align}
  &\p_t'\int(\rho_1v_{x0}+\rho_0v_{x1})\,\rmd z'+\p_X'\int(\rho_0v_{x0}v_{x0}-T_{xx0})\,\rmd z'=2\Omega\int(\rho_1v_{y0}+\rho_0v_{y1})\,\rmd z',\label{dintrhovx1}\\
  &\p_t'\int(\rho_1v_{y0}+\rho_0v_{y1})\,\rmd z'+\p_X'\int(\rho_0v_{y0}v_{x0}-T_{yx0})\,\rmd z'=-(2-q)\Omega\int(\rho_1v_{x0}+\rho_0v_{x1})\,\rmd z'.\label{dintrhovy1}
\end{align}
Finally, equation~(\ref{dintrhoz}) for $Z'$ at $O(\epsilon^0)$ and
$O(\epsilon^1)$ yields
\begin{align}
  &\p_t'\int\rho_0Z_0'\,\rmd z'=-\f{\rmi E_\nu}{\nu}\int\rho_0(\rmD_1\rmD_0+\rmD_0\rmD_1)\zeta_0\,\rmd z',\label{dintrhoz0}\\
  &\p_t'\int(\rho_1Z_0'+\rho_0Z_1')\,\rmd z'+\p_X'\int\left(\rho_0Z_0'v_{x0}-\f{\rmi T_{zx0}E_\nu}{\nu}\right)\,\rmd z'=-\f{\rmi E_\nu}{\nu}\int[\rho_1(\rmD_1\rmD_0+\rmD_0\rmD_1)+\rho_0(\rmD_2\rmD_0+\rmD_1\rmD_1+\rmD_0\rmD_2)]\zeta_0\,\rmd z',\label{dintrhoz1}
\end{align}
where
\begin{equation}
  Z'=Z_0'+\epsilon Z_1'+\cdots,\qquad
  Z_0'=\left(z'+\f{\rmi v_{z0}'}{\nu}\right)E_\nu,\qquad
  Z_1'=\f{\rmi v_{z1}'}{\nu}E_\nu.
\end{equation}

A sequence of deductions can be made from these equations. First, we
see from equation~(\ref{dintrho0}) that the surface density at leading
order,
\begin{equation}
  \Sigma_0(X,T)=\int\rho_0\,\rmd z',
\end{equation}
is independent of the fast time variable~$t$. This makes sense because
mass is conserved and the vertical oscillation does not cause any
horizontal mass transport. Second, equations (\ref{dintrhovx0}) and
(\ref{dintrhovy0}) imply that the mass-weighted mean horizontal
velocities at leading order undergo an unforced and undamped epicyclic
oscillation, because they can be combined into
\begin{equation}
  (\p_t'^2+\kappa^2)\int\rho_0v_{x0}\,\rmd z'=0,\qquad
  (\p_t'^2+\kappa^2)\int\rho_0v_{y0}\,\rmd z'=0.
\end{equation}
If the amplitude of this oscillation were non-zero, the disc could be
considered to have a non-zero eccentricity. Therefore we assume the
appropriate solution to be
\begin{equation}
  \int\rho_0v_{x0}\,\rmd z'=\int\rho_0v_{y0}\,\rmd z'=0.
\end{equation}
Third, we assume that the solution is periodic in the variable $t$
(i.e.\ purely oscillatory on the orbital timescale, as discussed
above), and carry out the averaging operation
\begin{equation}
  \langle\cdot\rangle=\f{1}{2\pi}\int_0^{2\pi}\cdot\,\rmd\tau
\end{equation}
on equations (\ref{dintrho2}), (\ref{dintrhovy1}) and
(\ref{dintrhoz1}) to eliminate some of the higher-order variables and
obtain
\begin{align}
  &\p_T\Sigma_0+\p_X(\Sigma_0\bar v_{x1})=0,\label{dsigma0}\\
  &\p_X\int\langle\rho_0v_{y0}v_{x0}-T_{yx0}\rangle\,\rmd z'=-(2-q)\Omega\Sigma_0\bar v_{x1},\label{massflux}\\
  &\p_X\int\left\langle\rho_0Z_0'v_{x0}-\f{\rmi T_{zx0}E_\nu}{\nu}\right\rangle\,\rmd z'=-\f{\rmi}{\nu}\int\left\langle E_\nu[\rho_1(\rmD_1\rmD_0+\rmD_0\rmD_1)+\rho_0(\rmD_2\rmD_0+\rmD_1\rmD_1+\rmD_0\rmD_2)]\zeta_0\right\rangle\rmd z',\label{dz}
\end{align}
where $\bar v_{x1}(X,T)$ is the mass-weighted mean radial velocity
defined by
\begin{equation}
  \Sigma_0\bar v_{x1}=\int\langle\rho_1v_{x0}+\rho_0v_{x1}\rangle\,\rmd z'.
\end{equation}
(We can safely write $\p_X'$ and $\p_T'$ as $\p_X$ and $\p_T$ here because they are acting on quantities that do not depend on $z'$.)
Equations (\ref{dintrhovx1}) and (\ref{dintrhoz0}) can also
be averaged in this way, but they do not yield any further information
that we require.

Equations (\ref{dsigma0}) and (\ref{massflux}) are related to those of
classical accretion-disc theory and can be combined into the
`diffusion equation'
\begin{equation}
  \p_T\Sigma_0=\f{1}{(2-q)\Omega}\p_X^2\int\langle\rho_0v_{y0}v_{x0}-T_{yx0}\rangle\,\rmd z'
\end{equation}
for the surface density.  (It has the character of a diffusion
equation if the stress integral on the right-hand side is a positive
and increasing function of the surface density.)  The evolution of the
surface density can be affected by the presence of a warp.

Equation~(\ref{dz}) determines the evolution of the warp. After some
integrations by parts on the right-hand side and use of the equation
of mass conservation at $O(\epsilon^1)$, we obtain
\begin{equation}
  \p_X\int\left\langle E_\nu\left[\rho_0z'v_{x0}-\f{\rmi(T_{zx0}-\rho_0v_{z0}'v_{x0})}{\nu}\right]\right\rangle\rmd z'=\left\langle\f{E_\nu}{\rmi\nu}\int\left[2\rho_0\zeta_{0tT}+(\rho_1v_{x0}+\rho_0v_{x1})(\p_t-\rmi\nu)\zeta_{0X}+\p_X'(\rho_0v_{x0}v_{x0}\zeta_{0X})\right]\rmd z'\right\rangle,
\end{equation}
which can be rearranged into the form
\begin{equation}
  \Sigma_0(Z_{0T}+\bar v_{x1}Z_{0X})+\p_X\int\left\langle E_\nu\left[\rho_0z'v_{x0}-\f{\rmi(T_{zx0}-\rho_0v_{z0}v_{x0})}{\nu}\right]\right\rangle\rmd z'=0.
\label{zt}
\end{equation}
The integral in equation~(\ref{zt}) is equivalent (apart from a factor
of $-\nu$) to the horizontal torque integral in equation~54 of
\citet{2013MNRAS.433.2403O}.

\subsubsection{Nonlinear oscillation equations}
\label{s:nonlinear}

Returning to the solution of the nonlinear oscillation equations
(\ref{dvx0})--(\ref{drho0}) in the case of a viscous disc, we first
rewrite them as
\begin{align}
  &\rmD_0v_{x0}-2\Omega v_{y0}=\f{1}{\rho_0}\p_z'(T_{xz0}+|\psi|\cos\tau\,T_{xx0}),\\
  &\rmD_0v_{y0}+(2-q)\Omega v_{x0}=\f{1}{\rho_0}\p_z'(T_{yz0}+|\psi|\cos\tau\,T_{yx0}),\\
  &\rmD_0v_{z0}=-\nu^2z'-\nu|\psi|\sin\tau\,v_{x0}+\f{1}{\rho_0}\p_z'(T_{zz0}+|\psi|\cos\tau\,T_{zx0}),\\
  &\rmD_0\rho_0=-\rho_0\Delta_0,
\end{align}
with
\begin{equation}
  \rmD_0=\nu\p_\tau+v_{z0}'\p_z',\qquad
  \Delta_0=\p_z'v_{z0}',\qquad
  v_{z0}'=v_{z0}+|\psi|\cos\tau\,v_{x0}.
\end{equation}
For a combination of isotropic pressure and viscous stress, we write
\begin{equation}
  T_{ij}=-p\,\delta_{ij}+\mu(\p_iu_j+\p_ju_i)+(\mu_\rmb-\twothirds\mu)(\p_ku_k)\delta_{ij},
\end{equation}
where $\mu$ and $\mu_\rmb$ are the dynamic shear and bulk
viscosities. The leading-order stress components that we require are
then
\begin{align}
  &T_{xx0}=-p_0+2\mu_0(|\psi|\cos\tau\p_z'v_{x0})+(\mu_{\rmb0}-\twothirds\mu_0)\Delta_0,\\
  &T_{xz0}=T_{zx0}=\mu_0(\nu|\psi|\sin\tau+|\psi|\cos\tau\,\p_z'v_{z0}+\p_z'v_{x0}),\\
  &T_{yx0}=\mu_0(-q\Omega+|\psi|\cos\tau\,\p_z'v_{y0}),\\
  &T_{yz0}=\mu_0(\p_z'v_{y0}),\\
  &T_{zz0}=-p_0+2\mu_0(\p_z'v_{z0})+(\mu_{\rmb0}-\twothirds\mu_0)\Delta_0.
\end{align}
For comparison with previous work, we simplify the thermal physics by considering an adiabatic flow, thereby neglecting viscous heating and radiative cooling. The differential identity
\begin{equation}
  \f{\rmd p}{\rho}=\rmd h-T\,\rmd s
\end{equation}
is used to rewrite pressure gradients in terms of gradients of the specific enthalpy $h$ and the specific entropy $s$ (in this equation only, $T$ denotes the temperature). The specific enthalpy and entropy of a perfect gas of adiabatic index $\gamma$ evolve according to
\begin{equation}
  \f{\rmD h}{\rmD t}=-(\gamma-1)h\Delta,\qquad
  \f{\rmD s}{\rmD t}=0,
\end{equation}
leading to
\begin{equation}
  \rmD_0h_0=-(\gamma-1)h_0\Delta_0,\qquad
  \rmD_0s_0=0.
\end{equation}
If we assume
\begin{equation}
  \mu_0=\f{\alpha p_0}{\nu},\qquad
  \mu_{\rmb0}=\f{\alpha_\rmb p_0}{\nu},
\end{equation}
where $\alpha$ and $\alpha_\rmb$ are dimensionless shear and bulk
viscosity coefficients that are independent of $z'$, then we can
separate the variables to write
\begin{equation}
  v_{x0}=u(\tau)\nu z',\qquad
  v_{y0}=v(\tau)\nu z',\qquad
  v_{z0}=w(\tau)\nu z',\qquad
  h_0=f(\tau)\nu^2-\f{1}{2}g(\tau)\nu^2z'^2,\qquad
  s_0=s(\tau),
\end{equation}
where $u$, $v$, $w$ and $g$ are dimensionless and satisfy the ordinary
differential equations
\begin{align}
  &\rmd_\tau u+(w+|\psi|\cos\tau\,u)u-2\beta v=|\psi|\cos\tau\,g-(\alpha_\rmb+\third\alpha)|\psi|\cos\tau\,g(w+|\psi|\cos\tau\,u)-\alpha g[|\psi|\sin\tau+(1+|\psi|^2\cos^2\tau)u],\\
  &\rmd_\tau v+(w+|\psi|\cos\tau\,u)v+(2-q)\beta u=-\alpha g[-q\beta|\psi|\cos\tau+(1+|\psi|^2\cos^2\tau)v],\\
  &\rmd_\tau w+(w+|\psi|\cos\tau\,u)w+|\psi|\sin\tau\,u=g-1-(\alpha_\rmb+\third\alpha)g(w+|\psi|\cos\tau\,u)-\alpha g[|\psi|^2\sin\tau\cos\tau+(1+|\psi|^2\cos^2\tau)w],\\
  &\rmd_\tau f=-(\gamma-1)(w+|\psi|\cos\tau\,u)f,\label{df}\\
  &\rmd_\tau g=-(\gamma+1)(w+|\psi|\cos\tau\,u)g,\label{dg}\\
  &\rmd_\tau s=0.
\end{align}
(Here we have suppressed the parametric dependence of the
solution on $X$ and $T$.)  These are exactly equivalent to equations
A37--A42 of \citet{2013MNRAS.433.2403O}, except for the inclusion
of the factor $\beta=\Omega/\nu$, which here could in principle differ
from unity.  In \citet{2013MNRAS.433.2403O} it is explained how these
equations are in turn exactly equivalent to equations 105--109 of
\citet{1999MNRAS.304..557O}.  (Note that these equations remain
invariant when $f$ is multiplied by a constant, so they do not fix the
normalization of $f$, which must instead be determined from the
surface density and entropy.)

The orbital stress averages we require are
\begin{align}
  &\int\langle\rho_0v_{y0}v_{x0}-T_{yx0}\rangle\,\rmd z'=-Q_1\nu^2\langle\mathcal{I}_0\rangle,\\
  &\int\left\langle E_\nu\left[\rho_0z'v_{x0}-\f{\rmi(T_{zx0}-\rho_0v_{z0}v_{x0})}{\nu}\right]\right\rangle\,\rmd z'=Q_4\psi\nu^2\langle\mathcal{I}_0\rangle,
\end{align}
where
\begin{equation}
  \mathcal{I}_0(X,T,t)=\int\rho_0z'^2\,\rmd z'
\end{equation}
is the second vertical moment of the density and $Q_1$ and
$Q_4=Q_2+\rmi Q_3$ are real and complex dimensionless coefficients
given by
\begin{equation}
  Q_1=\left\langle f_6[-uv+\alpha g(-q\beta+|\psi|\cos\tau\,v)]\right\rangle,\qquad
  Q_4|\psi|=\left\langle\rme^{\rmi\tau}f_6[u(1+\rmi w)-\rmi\alpha g(|\psi|\sin\tau+|\psi|\cos\tau\,w+u)]\right\rangle,
\end{equation}
where $f_6=fg^{-1}/\langle fg^{-1}\rangle$ describes the variation of
$\mathcal{I}_0$ with orbital phase. As explained in
\citet{2013MNRAS.433.2403O}, these expressions agree exactly with
equations~112 and~120 in \citet{1999MNRAS.304..557O}.

An issue not discussed in \citet{1999MNRAS.304..557O} is how to relate
$\langle\mathcal{I}_0\rangle$ to $\Sigma_0$ in a homentropic (or
polytropic) disc.  We may write
\begin{equation}
  \rho_0=Ch_0^n,
\end{equation}
where $C$ is a constant related to the specific entropy and $n$ is the
polytropic index given by $\gamma=1+1/n$.  (The dimensionless solution
given above has a specific entropy $s_0$ that is independent of $z'$,
but which could in principle depend on $X$ and $T$.  In that case we
should also solve an evolutionary equation for $s_0(X,T)$, which will
just be advected by the mean radial velocity $\bar v_{x1}$ in the
absence of non-adiabatic processes.)  We therefore have
\begin{equation}
  \Sigma_0=C(f\nu^2)^n\left(\f{2f}{g}\right)^{1/2}I_n,\qquad
  \mathcal{I}_0=C(f\nu^2)^n\left(\f{2f}{g}\right)^{3/2}J_n=\f{2f}{g}\f{\Sigma_0}{2n+3},
\end{equation}
where
\begin{equation}
  I_n=\int_{-1}^1(1-x^2)^n\,\rmd x=\f{\Gamma(n+1)\Gamma({\textstyle\f{1}{2}})}{\Gamma(n+{\textstyle\f{3}{2}})},\qquad
  J_n=\int_{-1}^1x^2(1-x^2)^n\,\rmd x=\f{I_n}{2n+3}
\end{equation}
are two dimensionless numbers.  The fact that $\Sigma_0$ is
independent of $\tau$ is related to the fact that $f^{\gamma+1}\propto
g^{\gamma-1}$, which can be seen from equations (\ref{df}) and
(\ref{dg}).  Eliminating $f$, we obtain
\begin{equation}
  \mathcal{I}_0=\f{\Sigma_0^{(3\gamma-1)/(\gamma+1)}g^{-2/(\gamma+1)}}{(n+{\textstyle\f{3}{2}})(CI_n\sqrt{2})^{2(\gamma-1)/(\gamma+1)}\nu^{4/(\gamma+1)}}.
\end{equation}
Thus
\begin{equation}
  \langle\mathcal{I}_0\rangle=C_\mathcal{I}Q_5\Sigma_0^{(3\gamma-1)/(\gamma+1)},
\end{equation}
where $C_\mathcal{I}$ is a dimensional constant that depends on the
entropy, and
\begin{equation}
  Q_5(|\psi|)=\left\langle g^{-2/(\gamma+1)}\right\rangle
\end{equation}
is a dimensionless function of the warp amplitude such that $Q_5(0)=1$.
This power-law relation between $\langle\mathcal{I}_0\rangle$ and
$\Sigma_0$, involving (for reasonable values of $\gamma$) a power
between $1$ and $2$ and a coefficient that depends on the warp
amplitude, is similar to what is obtained for a radiative disc
\citep{2000MNRAS.317..607O}.

In the isothermal case $\gamma=1$, we have instead $f_6=g^{-1}/\langle g^{-1}\rangle$ (because $f$ becomes a constant in the limit $\gamma\to1$) and
\begin{equation}
  \langle\mathcal{I}_0\rangle=c_\rms^2Q_5\Sigma_0,\qquad
  Q_5(|\psi|)=\left\langle g^{-1}\right\rangle.
\end{equation}
When $Q_1$ and $Q_4$ are combined with $Q_5$, the factor
$\langle g^{-1}\rangle$ cancels out, leaving the expressions 91 and 92
in \citet{2013MNRAS.433.2403O}.  In the interests of consistency
between the isothermal and polytropic cases, these expressions should
really be regarded as definitions of $Q_1Q_5$ and $Q_4Q_5$ rather
than $Q_1$ and $Q_4$, and we adopt this convention henceforth.

\subsubsection{Summary of the non-resonant case}
\label{s:summary}

When we remove the asymptotic scalings and subscripts, the
evolutionary equations we have derived take the form
\begin{align}
  &\f{\p\Sigma}{\p t}+\f{\p(\Sigma\bar v)}{\p x}=0,\\
  &\Sigma\left(\f{\p Z}{\p t}+\bar v\f{\p Z}{\p x}\right)=\f{\p}{\p x}\left(Q_4\nu^2\langle\mathcal{I}\rangle\f{\p Z}{\p x}\right),
\end{align}
together with
\begin{align}
  &(2-q)\Omega\Sigma\bar v=\f{\p}{\p x}\left(Q_1\nu^2\langle\mathcal{I}\rangle\right),\\
  &\langle\mathcal{I}\rangle=C_\mathcal{I}Q_5\Sigma^{(3\gamma-1)/(\gamma+1)},
\end{align}
where $Q_1$ (real), $Q_4$ (complex) and $Q_5$ (real, positive) are
nonlinear functions of
\begin{equation}
  |\psi|=\left|\f{\p Z}{\p x}\right|.
\end{equation}

Overall we obtain a system of equations for the evolution of the
surface density $\Sigma(x,t)$ and the vertical amplitude $Z(x,t)$ that
are very similar to those of \citet{1999MNRAS.304..557O} for warped
discs, with $Z$ playing the role of $-r(l_x+\rmi l_y)$, and with
exactly the same coefficients $Q_i$. The only
differences are that certain factors of $r$ coming from the global,
spherical geometry do not appear in the local model, and that terms
involving $Q_2|\psi|^2$ are absent. Our equations are a consistent
simplification of those of \citet{1999MNRAS.304..557O} for a warp that
varies on a lengthscale that is small compared to $r$; this makes
sense because we derived them in a local approximation.

The local model admits a special solution in the form of a uniformly
travelling (and generally decaying) bending wave,
\begin{equation}
  Z=A\,\rme^{\rmi kx},\qquad
  \Sigma=\cst,\qquad
  \bar v=0,
\end{equation}
where $A(t)$ is a complex amplitude and $k$ is a constant real
wavenumber.  This solution has
\begin{equation}
  \psi=-Z_x=-\rmi kA\,\rme^{\rmi kx},\qquad
  |\psi|=|kA|,
\end{equation}
so that $|\psi|$ and the coefficients $Q_i$ are independent of $x$.
It represents a twisted warp of uniform amplitude.  The evolutionary
equation for $Z$ reduces to the first-order ordinary differential
equation
\begin{equation}
  \dot A=-\f{Q_4\nu^2\langle\mathcal{I}\rangle k^2}{\Sigma}A.
\end{equation}
The warp amplitude therefore decays according to
\begin{equation}
  \f{\rmd}{\rmd t}(\ln|\psi|)=-Q_2Q_5\nu^2C_\mathcal{I}\Sigma^{2(\gamma-1)/(\gamma+1)}k^2,
\end{equation}
while the phase evolves according to
\begin{equation}
  \f{\rmd}{\rmd t}(\arg\psi)=-Q_3Q_5\nu^2C_\mathcal{I}\Sigma^{2(\gamma-1)/(\gamma+1)}k^2.
\end{equation}
This pair of equations can be thought of as a nonlinear dispersion
relation showing how the decay rate and angular frequency of a
travelling wave depend on its wavenumber and amplitude.  The decay is
not exactly exponential because the decay rate depends on amplitude
through the function $Q_2(|\psi|)Q_5(|\psi|)$.

It is known that the nonlinear diffusion of warps can be subject to an instability, which may cause a warp to steepen into a break \citep{2000MNRAS.317..607O,2018MNRAS.476.1519D,2020MNRAS.495.1148D,2021ApJ...909...81R}. The instability results from the dependence of the coefficients $Q_i$ on $|\psi|$. The special solution described above has a uniform warp amplitude $|\psi|$ and therefore does not exhibit this behaviour, but it could be linearly unstable to perturbations that modulate the warp amplitude.

\subsection{Resonant case}

The analysis of Section~\ref{s:non-resonant} breaks down in the
resonant case $\kappa=\nu$ if the disc has only an isotropic stress
from gas pressure.  In this case the forcing of horizontal
oscillations by the slowly modulated corrugation is resonant and
undamped.  An alternative asymptotic scaling that works in this case
is
\begin{align}
  &x=\epsilon^{-1}X,\qquad
  t=\epsilon^{-1}T,\\
  &\zeta=\zeta_0(X,t,T)=\mathrm{Re}\left[Z_0(X,T)\,\rme^{-\rmi\nu t}\right],\\
  &v_x=v_{x0}(X,z',t,T)+\epsilon v_{x1}(X,z',t,T)+\cdots,\\
  &v_y=v_{y0}(X,z',t,T)+\epsilon v_{y1}(X,z',t,T)+\cdots,\\
  &v_z'=\epsilon v_{z1}'(X,z',t,T)+\cdots,\\
  &\rho=\rho_0(z')+\epsilon\rho_1(X,z',t,T)+\cdots,\\
  &p=p_0(z')+\epsilon p_1(X,z',t,T)+\cdots,
\end{align}
so that
\begin{equation}
  \rmD=\rmD_0+\epsilon\rmD_1+\epsilon^2\rmD_2+\cdots,
\end{equation}
with
\begin{equation}
  \rmD_0=\p_t',\qquad
  \rmD_1=\p_T'+v_{x0}\p_X'+v_{z1}'\p_z',\qquad
  \rmD_2=v_{x1}\p_X'+v_{z2}'\p_z'.
\end{equation}
The meaning of these expressions is somewhat different from that of
the non-resonant case.  The corrugation is now of comparable amplitude
to the vertical scaleheight.  It consists of a harmonic oscillation on
the orbital timescale, with angular frequency $\nu$, and with an
amplitude and phase (described by the complex amplitude $Z_0$) that
vary on a radial lengthscale that is long ($\epsilon^{-1}$) compared
to the vertical scaleheight and on a timescale that is slow
($\epsilon^{-1}$) compared to the orbital timescale.  Despite the
reduced amplitude of the corrugation, the horizontal internal
velocities are still comparable to the sound speed because they are
driven at resonance.  Apart from being translated by the vertical
oscillation, the density and pressure experience relatively small
($\epsilon$) fractional perturbations.

We substitute these expansions again into the basic equations
(\ref{dvx})--(\ref{drho}) and compare terms of the same order in
$\epsilon$. The vertical component~(\ref{dvz}) of the equation of
motion at $O(\epsilon^0)$ yields
\begin{equation}
  0=-\nu^2z'-(\p_t'^2+\nu^2)\zeta_0-\f{1}{\rho_0}\p_z'p_0.
\end{equation}
The assumed form of $\zeta_0$ means that $(\p_t'^2+\nu^2)\zeta_0$
vanishes, leaving the standard equation of vertical hydrostatic
equilibrium,
\begin{equation}
  0=-\nu^2z'-\f{1}{\rho_0}\p_z'p_0,
\end{equation}
involving quantities that depend only on $z'$.

The horizontal components (\ref{dvx}) and (\ref{dvy}) at $O(\epsilon^0)$ give
\begin{equation}
  \p_t'v_{x0}-2\Omega v_{y0}=0,\qquad
  \p_t'v_{y0}+\f{\kappa^2}{2\Omega}v_{x0}=0,
\end{equation}
which admit a free epicyclic motion of the form
\begin{equation}
  v_{x0}=\mathrm{Re}\left[U(X,z',T)\,\rme^{-\rmi\kappa t}\right],\qquad
  v_{y0}=\mathrm{Re}\left[-\f{\rmi\kappa}{2\Omega}\,U(X,z',T)\,\rme^{-\rmi\kappa t}\right],
\end{equation}
where $U$ is a complex amplitude to be determined subsequently. The
epicyclic motion appears to be free at this stage because it is in
fact forced resonantly.

At $O(\epsilon^1)$ equations (\ref{dvx})--(\ref{drho}) yield (using
the hydrostatic balance)
\begin{align}
  &\rmD_1v_{x0}+\p_t'v_{x1}-2\Omega v_{y1}=-\nu^2z'\zeta_{0X},\label{dvx1}\\
  &\rmD_1v_{y0}+\p_t'v_{y1}+\f{\kappa^2}{2\Omega}v_{x1}=0,\label{dvy1}\\
  &\p_t'v_{z1}'=-(\rmD_1\rmD_0+\rmD_0\rmD_1)\zeta_0-\f{\rho_1}{\rho_0}\nu^2z'-\f{1}{\rho_0}\p_z'p_1,\label{dvz1}\\
  &v_{z1}'\p_z'\rho_0+\p_t'\rho_1=-\rho_0(\p_X'v_{x0}+\p_z'v_{z1}').\label{drho1}
\end{align}
For adiabatic flow we also have a corresponding equation for the
pressure,
\begin{equation}
  v_{z1}'\p_z'p_0+\p_t'p_1=-\gamma p_0(\p_X'v_{x0}+\p_z'v_{z1}'),
\label{dp1}
\end{equation}
where $\gamma$ is the adiabatic index.  Note that
\begin{equation}
  (\rmD_1\rmD_0+\rmD_0\rmD_1)\zeta_0=\zeta_{0X}\p_t'v_{x0}+2(\zeta_{0tT}+v_{x0}\zeta_{0Xt}).
\end{equation}

Equations (\ref{dvx1}) and (\ref{dvy1}) can be combined into
\begin{equation}
  (\p_t'^2+\kappa^2)v_{x1}=F_{\rmh1},
\label{fh1}
\end{equation}
with horizontal forcing
\begin{equation}
  F_{\rmh1}=-\nu^2z'\zeta_{0Xt}-\p_t'\rmD_1v_{x0}-2\Omega\rmD_1v_{y0},
\end{equation}
which evaluates to
\begin{equation}
  F_{\rmh1}=\mathrm{Re}\left\{\rmi\nu^3z'Z_{0X}\,\rme^{-\rmi\nu t}+[2\rmi\kappa U_T-U_{z'}(\p_t'-2\rmi\kappa)v_{z1}']\,\rme^{-\rmi\kappa t}+\f{\rmi\kappa}{2}U^*U_X+\f{3\rmi\kappa}{2}UU_X\,\rme^{-2\rmi\kappa t}\right\}.
\end{equation}
The linear operator on the left-hand side of equation~(\ref{fh1}) is
self-adjoint and has null eigenfunctions $\rme^{\mp\rmi\kappa t}$
representing free epicyclic oscillations with an arbitrary vertical
structure.  The corresponding solvability conditions are
\begin{equation}
  \int F_{\rmh1}\,\rme^{\pm\rmi\kappa t}\,\rmd t=0,
\end{equation}
where the integration is over one period of the epicyclic oscillation.
Given that $\kappa=\nu$, the first term in $F_{\rmh1}$, which is the
forcing of the epicyclic oscillations by the warp, is resonant and
contributes to the solvability conditions, which become
\begin{equation}
  U_T=-\f{1}{2}\nu^2z'Z_{0X}
\label{sch}
\end{equation}
and the complex conjugate of this equation.  The term
$-U_{z'}(\p_t'-2\rmi\kappa)v_{z1}'$ cannot contribute to the
solvability conditions because to do so $v_{z1}'$ would need to
contain terms proportional to either $\rme^{2\rmi\kappa t}$ or
$\rme^{0t}$. In the first case the term vanishes on application of
$(\p_t'-2\rmi\kappa)$. In the second case there would have to be a
non-zero mean relative vertical velocity, which we exclude in the next
paragraph.

Using the hydrostatic condition, equations (\ref{dvz1})--(\ref{dp1})
can be combined into
\begin{equation}
  -\p_z'(\gamma p_0\p_z'v_{z1}')+\rho_0(\p_t'^2+\nu^2)v_{z1}'=F_{\rmv1},
\label{fv1}
\end{equation}
with vertical forcing
\begin{equation}
  F_{\rmv1}=\p_z'(\gamma p_0\p_X'v_{x0})+\rho_0\nu^2z'\p_X'v_{x0}-\rho_0\p_t'(\rmD_1\rmD_0+\rmD_0\rmD_1)\zeta_0.
\end{equation}
Given that $\kappa=\nu$, this evaluates to
\begin{equation}
  F_{\rmv1}=\mathrm{Re}\left\{[\p_z'(\gamma p_0U_X)+\rho_0\nu^2(z'U_X+2Z_{0T})]\,\rme^{-\rmi\nu t}+3\rho_0\nu^2 UZ_{0X}\,\rme^{-2\rmi\nu t}\right\}.
\end{equation}
So there are no non-oscillatory contributions to $F_{\rmv1}$ or to
$v_{z1}'$. This justifies the step taken above in deriving
equation~(\ref{sch}).

The linear operator on the left-hand side of equation~(\ref{fv1}) is
also self-adjoint and has null eigenfunctions $\rme^{\mp\rmi\nu
  t}$ representing free vertical oscillations independent of $z'$.
The corresponding solvability conditions are
\begin{equation}
  \iint F_{\rmv1}\,\rme^{\pm\rmi\nu t}\,\rmd z'\,\rmd t=0,
\end{equation}
where the integration is over the full vertical extent of the disc and
over one period of the vertical oscillation.  Thus we obtain
\begin{equation}
  \int\rho_0Z_{0T}\,\rmd z'=-\f{1}{2}\int\rho_0z'U_X\,\rmd z'
\label{scv}
\end{equation}
and the complex conjugate of this equation.  Combining this with
equation~(\ref{sch}), we obtain the wave equation
\begin{equation}
  Z_{0TT}=\f{1}{4}\nu^2H^2Z_{0XX},
\end{equation}
where the scaleheight $H$ is defined by
\begin{equation}
  \int\rho_0z'^2\,\rmd z'=H^2\int\rho_0\,\rmd z'.
\end{equation}
When the asymptotic scalings are removed, the equation takes the form
\begin{equation}
  \f{\p^2Z}{\p t^2}=\f{1}{4}\nu^2H^2\f{\p^2Z}{\p x^2}.
\end{equation}
This result shows that the warp propagates in the form of
non-dispersive waves, with wave speeds $\pm\half\nu H$.  In the case of
a spherically symmetric potential with $\nu=\Omega$, this agrees with
the result obtained in cylindrical geometry by
\citet{1995ApJ...438..841P}. It also agrees with the dipsersion relation~(\ref{resonant}).

\section{Computational considerations}
\label{s:computational}

We have made some preliminary numerical investigations of the local model for warped discs by using the PLUTO \citep{2007ApJS..170..228M} and Athena++ \citep{2020ApJS..249....4S} codes to solve the equations
of ideal gas dynamics for an isothermal gas in a 2D Cartesian domain ($x$ and $z$ coordinates, but with
three velocity components) using a finite-volume method. A uniform kinematic viscosity can also be included. Equations
(\ref{ddvx})--(\ref{isothermal}) can be solved using the standard shearing-box modules of these codes. The use of periodic boundary conditions in the $x$ direction allows propagating bending waves and warps to be studied without end-effects. Provided that the domain is sufficiently long in the $x$ direction, it is possible to access the regime of astrophysical interest, in which the radial wavelength of the warp is long compared to the vertical scaleheight of the disc.

These preliminary investigations, which we do not report in detail here, confirm that the dynamics of warped discs and the propagation of bending waves can be studied using this model, and it is possible to observe the occurrence of the parametric instability \citep{2000MNRAS.318.1005G,2013MNRAS.433.2420O,2019MNRAS.483.3738P}. We note here some of the considerations that will be important for a more detailed computational study.

The height at which the vertical boundaries are placed, and the nature of those boundaries, can have an important effect on the outcome.  None of the standard boundary conditions (periodic, reflecting or outflow) is well suited to the desired solution in which the gas oscillates freely through the boundary.  The vertical motion can be transonic if the amplitude is sufficiently large, and the presence of boundaries causes shocks.  The damping of the oscillatory vertical motion resulting from these shocks depends on the location and nature of the boundaries and needs to be quantified.
A Lagrangian method that can follow the free vertical oscillation of the disc would have a clear advantage here.  However, for subsonic vertical motion and reflecting boundary conditions at several scaleheights from the midplane, we found that the damping was very small.

It can be useful to view the simulations stroboscopically, once per orbit.  This method filters out the basic vertical oscillation and reveals the modulatory dynamics that corresponds to the slow evolution of the warp in the non-rotating frame.

Although the periodic radial boundaries are artificial, and mean that a propagating warp cycles through the domain, they have the advantage of being compatible with special solutions such as the twisted warp of uniform amplitude, discussed in Section~\ref{s:summary}.

It may be useful to compare the approach proposed in this paper, which represents a warp within a standard shearing box (SSB), with that of the warped shearing box (WSB) defined by \citet{2013MNRAS.433.2403O} and used in nonlinear hydrodynamic simulations by \citet{2019MNRAS.483.3738P}. In the SSB the warp is represented explicitly and evolves freely as a result of the dynamics occurring within the box, whereas in the WSB a warp of fixed amplitude is imposed through the oscillatory coordinate system, and the evolution of the warp is to be deduced from the torques measured in the box. The SSB should be much larger in the radial ($x$) direction to incorporate the scale of the warp explicitly, whereas the WSB can zoom in to a region that is small compared to the scale of the warp. Simulations in the SSB can make use of existing publicly available codes, while the WSB requires the coding of a novel set of equations. Finally, the SSB has to deal with oscillatory flows through the vertical boundaries (as discussed above), while the WSB naturally follows this motion, although in nonlinear warp regimes it may still have to deal with strong vertical compressions of the disc.

\section{Conclusion}
\label{s:conclusion}

In this paper we have shown that many aspects of the dynamics of
warped discs can be studied in the local approximation, which is the
basis for the well known model of the shearing box.  We have
demonstrated that the warping of a disc corresponds, in the local
model, to a locally axisymmetric corrugation of the midplane of the
disc that oscillates vertically at the orbital frequency, while
evolution of the warp corresponds to a modulation of the complex
amplitude of the vertical oscillation.  We have derived a conservation
law for this amplitude that is the local equivalent of the
conservation of angular momentum.  For lengthscales that are long
compared to the vertical scaleheight, the non-resonant and resonant
regimes of warp dynamics, including the diffusive and wavelike regimes
of Keplerian discs, occur in the local model in the same way as in the
global model. This opens the possibility of studying the local physics of warped discs at high resolution using standard computational methods.

\section*{Acknowledgements}

This research was supported by STFC through grants 
ST/P000673/1 and ST/T00049X/1. I thank the referee for a very careful reading of the manuscript and for suggestions that led to its improvement.

\section*{Data availability}

No new data were generated or analysed in support of this research.

\bsp	
\label{lastpage}

\begin{thebibliography}{}

\bibitem[\protect\citeauthoryear{Bardeen \& Petterson}{1975}]{1975ApJ...195L..65B} Bardeen J.~M., Petterson J.~A., 1975, ApJL, 195, L65. doi:10.1086/181711

\bibitem[\protect\citeauthoryear{Bender
\& Orszag}{1978}]{BO78} Bender C.~M., Orszag S.~A., 1978, Advanced Mathematical Methods for Scientists and Engineers I: Asymptotic Methods and Perturbation Theory, McGraw-Hill

\bibitem[\protect\citeauthoryear{Bohn et al.}{2022}]{2022A&A...658A.183B} Bohn A.~J., Benisty M., Perraut K., van der Marel N., W{\"o}lfer L., van Dishoeck E.~F., Facchini S., et al., 2022, A\&A, 658, A183. doi:10.1051/0004-6361/202142070

\bibitem[\protect\citeauthoryear{Casassus et al.}{2019}]{2019MNRAS.486L..58C} Casassus S., P{\'e}rez S., Osses A., Marino S., 2019, MNRAS, 486, L58. doi:10.1093/mnrasl/slz059

\bibitem[\protect\citeauthoryear{Do{\v{g}}an et al.}{2018}]{2018MNRAS.476.1519D} Do{\v{g}}an S., Nixon C.~J., King A.~R., Pringle J.~E., 2018, MNRAS, 476, 1519. doi:10.1093/mnras/sty155

\bibitem[\protect\citeauthoryear{Do{\u{g}}an \& Nixon}{2020}]{2020MNRAS.495.1148D} Do{\u{g}}an S., Nixon C.~J., 2020, MNRAS, 495, 1148. doi:10.1093/mnras/staa1239

\bibitem[\protect\citeauthoryear{Gammie, Goodman \& Ogilvie}{2000}]{2000MNRAS.318.1005G} Gammie C.~F., Goodman J., Ogilvie G.~I., 2000, MNRAS, 318, 1005. doi:10.1046/j.1365-8711.2000.03669.x

\bibitem[\protect\citeauthoryear{Gerend \& Boynton}{1976}]{1976ApJ...209..562G} Gerend D., Boynton P.~E., 1976, ApJ, 209, 562. doi:10.1086/154751

\bibitem[\protect\citeauthoryear{Hatchett, Begelman \& Sarazin}{1981}]{1981ApJ...247..677H} Hatchett S.~P., Begelman M.~C., Sarazin C.~L., 1981, ApJ, 247, 677. doi:10.1086/159079

\bibitem[\protect\citeauthoryear{Hawley, Gammie \& Balbus}{1995}]{1995ApJ...440..742H} Hawley J.~F., Gammie C.~F., Balbus S.~A., 1995, ApJ, 440, 742. doi:10.1086/175311

\bibitem[\protect\citeauthoryear{Hawley \& Krolik}{2019}]{2019ApJ...878..149H} Hawley J.~F., Krolik J.~H., 2019, ApJ, 878, 149. doi:10.3847/1538-4357/ab1f6e

\bibitem[\protect\citeauthoryear{Kumar \& Pringle}{1985}]{1985MNRAS.213..435K} Kumar S., Pringle J.~E., 1985, MNRAS, 213, 435. doi:10.1093/mnras/213.3.435

\bibitem[\protect\citeauthoryear{Lai}{1999}]{1999ApJ...524.1030L} Lai D., 1999, ApJ, 524, 1030. doi:10.1086/307850

\bibitem[\protect\citeauthoryear{Larwood et al.}{1996}]{1996MNRAS.282..597L} Larwood J.~D., Nelson R.~P., Papaloizou J.~C.~B., Terquem C., 1996, MNRAS, 282, 597. doi:10.1093/mnras/282.2.597

\bibitem[\protect\citeauthoryear{Liska et al.}{2019}]{2019MNRAS.487..550L} Liska M., Tchekhovskoy A., Ingram A., van der Klis M., 2019, MNRAS, 487, 550. doi:10.1093/mnras/stz834

\bibitem[\protect\citeauthoryear{Lubow}{1992}]{1992ApJ...398..525L} Lubow S.~H., 1992, ApJ, 398, 525. doi:10.1086/171877

\bibitem[\protect\citeauthoryear{Lubow \& Ogilvie}{2000}]{2000ApJ...538..326L} Lubow S.~H., Ogilvie G.~I., 2000, ApJ, 538, 326. doi:10.1086/309101

\bibitem[\protect\citeauthoryear{Mignone et al.}{2007}]{2007ApJS..170..228M} Mignone A., Bodo G., Massaglia S., Matsakos T., Tesileanu O., Zanni C., Ferrari A., 2007, ApJS, 170, 228. doi:10.1086/513316

\bibitem[\protect\citeauthoryear{Miyoshi et al.}{1995}]{1995Natur.373..127M} Miyoshi M., Moran J., Herrnstein J., Greenhill L., Nakai N., Diamond P., Inoue M., 1995, Natur, 373, 127. doi:10.1038/373127a0

\bibitem[\protect\citeauthoryear{Nelson \& Papaloizou}{1999}]{1999MNRAS.309..929N} Nelson R.~P., Papaloizou J.~C.~B., 1999, MNRAS, 309, 929. doi:10.1046/j.1365-8711.1999.02894.x

\bibitem[\protect\citeauthoryear{Ogilvie}{1999}]{1999MNRAS.304..557O} Ogilvie G.~I., 1999, MNRAS, 304, 557. doi:10.1046/j.1365-8711.1999.02340.x

\bibitem[\protect\citeauthoryear{Ogilvie}{2000}]{2000MNRAS.317..607O} Ogilvie G.~I., 2000, MNRAS, 317, 607. doi:10.1046/j.1365-8711.2000.03654.x

\bibitem[\protect\citeauthoryear{Ogilvie}{2006}]{2006MNRAS.365..977O} Ogilvie G.~I., 2006, MNRAS, 365, 977. doi:10.1111/j.1365-2966.2005.09776.x

\bibitem[\protect\citeauthoryear{Ogilvie \& Latter}{2013a}]{2013MNRAS.433.2403O} Ogilvie G.~I., Latter H.~N., 2013, MNRAS, 433, 2403. doi:10.1093/mnras/stt916

\bibitem[\protect\citeauthoryear{Ogilvie \& Latter}{2013b}]{2013MNRAS.433.2420O} Ogilvie G.~I., Latter H.~N., 2013, MNRAS, 433, 2420. doi:10.1093/mnras/stt917

\bibitem[\protect\citeauthoryear{Ogilvie \& Lubow}{1999}]{1999ApJ...515..767O} Ogilvie G.~I., Lubow S.~H., 1999, ApJ, 515, 767. doi:10.1086/307037

\bibitem[\protect\citeauthoryear{Okazaki, Kato, \& Fukue}{1987}]{1987PASJ...39..457O} Okazaki A.~T., Kato S., Fukue J., 1987, PASJ, 39, 457

\bibitem[\protect\citeauthoryear{Paardekooper \& Ogilvie}{2019}]{2019MNRAS.483.3738P} Paardekooper S.-J., Ogilvie G.~I., 2019, MNRAS, 483, 3738. doi:10.1093/mnras/sty3349

\bibitem[\protect\citeauthoryear{Papaloizou \& Lin}{1995}]{1995ApJ...438..841P} Papaloizou J.~C.~B., Lin D.~N.~C., 1995, ApJ, 438, 841. doi:10.1086/175127

\bibitem[\protect\citeauthoryear{Papaloizou \& Pringle}{1983}]{1983MNRAS.202.1181P} Papaloizou J.~C.~B., Pringle J.~E., 1983, MNRAS, 202, 1181. doi:10.1093/mnras/202.4.1181

\bibitem[\protect\citeauthoryear{Papaloizou \& Terquem}{1995}]{1995MNRAS.274..987P} Papaloizou J.~C.~B., Terquem C., 1995, MNRAS, 274, 987. doi:10.1093/mnras/274.4.987

\bibitem[\protect\citeauthoryear{Paris \& Ogilvie}{2018}]{2018MNRAS.477.2406P} Paris J.~B., Ogilvie G.~I., 2018, MNRAS, 477, 2406. doi:10.1093/mnras/sty596

\bibitem[\protect\citeauthoryear{Petterson}{1978}]{1978ApJ...226..253P} Petterson J.~A., 1978, ApJ, 226, 253. doi:10.1086/156604

\bibitem[\protect\citeauthoryear{Pringle}{1992}]{1992MNRAS.258..811P} Pringle J.~E., 1992, MNRAS, 258, 811. doi:10.1093/mnras/258.4.811

\bibitem[\protect\citeauthoryear{Pringle}{1996}]{1996MNRAS.281..357P} Pringle J.~E., 1996, MNRAS, 281, 357. doi:10.1093/mnras/281.1.357

\bibitem[\protect\citeauthoryear{Raj, Nixon \& Do{\u{g}}an}{2021}]{2021ApJ...909...81R} Raj A., Nixon C.~J., Do{\u{g}}an S., 2021, ApJ, 909, 81. doi:10.3847/1538-4357/abdc24

\bibitem[\protect\citeauthoryear{Sakai et al.}{2019}]{2019Natur.565..206S} Sakai N., Hanawa T., Zhang Y., Higuchi A.~E., Ohashi S., Oya Y., Yamamoto S., 2019, Natur, 565, 206. doi:10.1038/s41586-018-0819-2

\bibitem[\protect\citeauthoryear{Skowron et al.}{2019}]{2019Sci...365..478S} Skowron D.~M., Skowron J., Mr{\'o}z P., Udalski A., Pietrukowicz P., Soszy{\'n}ski I., Szyma{\'n}ski M.~K., et al., 2019, Sci, 365, 478. doi:10.1126/science.aau3181

\bibitem[\protect\citeauthoryear{Stone et al.}{2020}]{2020ApJS..249....4S} Stone J.~M., Tomida K., White C.~J., Felker K.~G., 2020, ApJS, 249, 4. doi:10.3847/1538-4365/ab929b

\bibitem[\protect\citeauthoryear{Xiang-Gruess \& Papaloizou}{2013}]{2013MNRAS.431.1320X} Xiang-Gruess M., Papaloizou J.~C.~B., 2013, MNRAS, 431, 1320. doi:10.1093/mnras/stt254

\end{thebibliography}
\end{document}